# Reversible Tuning of the Collapsed Tetragonal Phase Transition in $CaFe_2As_2$ by separate control of chemical pressure and electron doping

K. Zhao[1], C. Stingl[1], R. S. Manna[1], C.Q. Jin[2,3], and P. Gegenwart[1]

1) *Experimentalphysik VI, Center for Electronic Correlations and Magnetism, Augsburg University, 86159 Augsburg, Germany*
2) *Beijing National Laboratory for Condensed Matter Physics, and Institute of Physics, Chinese Academy of Sciences, Beijing 100190, China*
3) *Collaborative Innovation Center of Quantum Matter, Beijing, China*

## Abstract

Single crystals of $Ca(Fe_{1-x}Ru_x)_2As_2$ ($0 \leq x \leq 0.065$) and $Ca_{1-y}La_y(Fe_{0.973}Ru_{0.027})_2As_2$ ($0 \leq y \leq 0.2$) have been synthesized and studied with respect to their structural, electronic and magnetic properties. The partial substitution of Fe by Ru induces a decrease of the c-axis constant leading for $x \leq 0.023$ to a suppression of the coupled magnetic and structural (tetragonal to orthorhombic) transitions. At $x_{cr}=0.023$ a first order transition to a collapsed tetragonal (CT) phase is found, which behaves like a Fermi liquid and which is stabilized by further increase of x. The absence of superconductivity near $x_{cr}$ is consistent with truly hydrostatic pressure experiments on undoped $CaFe_2As_2$. Starting in the CT regime at $x=0.027$ we investigate the additional effect of electron doping by partial replacement of Ca by La. Most remarkably, with increasing y the CT phase transition is destabilized and the system is tuned back into a tetragonal ground state at $y \geq 0.08$. This effect is ascribed to a weakening of interlayer As-As bonds by electron doping. Upon further electron doping filamentary superconductivity with $T_c$ of 41 K at $y=0.2$ is observed.

# Introduction

Since the discovery of the iron arsenide based superconductor $LaO_{1-x}F_xFeAs$ with superconducting transition temperature $T_c$=26 K,[1] a series of "1111", "122", "111" or "11" type pnictide superconductors, together with chalcogenide superconductors were synthesized.[2-9] The $AFe_2As_2$ (A = Ca, Sr, Ba, or Eu) "122" parent compounds display structural and magnetic transitions from a high-temperature tetragonal (Tet) paramagnetic to a low-temperature orthorhombic (Ort) magnetic phase. Pressure and chemical doping have similar effects on tuning the electric structure and phase diagram of $AFe_2As_2$ systems.[10] $CaFe_2As_2$ is special among the 122 materials as due to its reduced unit cell volume and c-lattice parameter, it is very close to collapsed tetragonal (CT) phase. At ambient conditions, $CaFe_2As_2$ displays a first-order combined Tet to Ort and magnetic transition at 165 K.[11, 12] It is extremely sensitive to the pressure conditions, including both the pressure cell type and the pressure transmitting medium.[13] $CaFe_2As_2$ has been tuned to a collapsed tetragonal phase at pressures above 0.35 GPa at 50 K[14, 15], with a pressure induced superconducting phase between 0.23 and 0.86 GPa detected in a nonmagnetic, piston-cylinder-type, Be-Cu pressure cell.[16] However, no superconductivity is observed for single crystal $CaFe_2As_2$ under truly hydrostatic conditions in a helium gas pressure system.[17]

The CT phase has also been induced by strain[18] or chemical substitution in $CaFe_2(As_{1-x}P_x)_2$[19], $Ca(Fe_{1-x}Rh_x)_2As_2$[20] and $Ca_{1-x}Pr_xFe_2As_2$[21]. In the former two materials, The CT phase transition results in large c-parameter changes and leads to a strong change of the electrical resistivity and the magnetic properties, which indicate ordinary Fermi liquid behavior. For $Ca_{1-x}Pr_xFe_2As_2$ the electronic structure within the CT phase is different compared to $CaFe_2As_2$ under pressure (or chemical pressure), likely due to the charge carrier doping effect by $Pr^{3+}$.[22-24] Because of its significantly smaller ionic radius compared to Ca, Pr substitution introduces both chemical pressure and electron doping.[21] It is therefore highly interesting to disentangle the influence of these two effects.

Previously, it has been demonstrated for $Ba(Fe_{1-x}Ru_x)_2As_2$, that the isovalent substitution of Fe by Ru effectively acts as chemical pressure. Although Ru is bigger than Fe, the c-parameter and c/a ratio shrink in this series and the phase diagram versus x could be scaled by pressure: both 3 GPa of applied pressure or 10% Ru substitution for Fe affects the phase diagram in a similar manner.[25, 26] However, up to now the effect of Ru doping in $Ca(Fe_{1-x}Ru_x)_2As_2$ has not been investigated. We therefore decided to first investigate the effect of chemical pressure in $Ca(Fe_{1-x}Ru_x)_2As_2$ and subsequently to study the influence of electron doping by co-doping of divalent Ca by trivalent La in $Ca_{1-y}La_y(Fe_{1-x}Ru_x)_2As_2$ for selected chemical pressure. Note, that due to the similar ionic radii of La and Ca, the electron doping by La substitution does not induce additional chemical pressure. We can therefore systematically tune the $CaFe_2As_2$ ground state by chemical pressure into the CT phase and subsequently investigate the influence of electron doping within this state. Most interestingly, electron doping destabilizes the CT phase and allows for the first-time to tune the ground state back towards the tetragonal phase. We also demonstrate the emergence of superconductivity with $T_c$ above 40 K when the CT phase is suppressed by electron doping.

After a short summary of experimental details we first discuss the temperature-composition (T−x) phase diagram of $Ca(Fe_{1-x}Ru_x)_2As_2$ based on structural, magnetic, and transport measurements. We establish the emergence of the CT ground state at x>0.023 and further systematically investigate the influence of electron doping in $Ca_{1-y}La_y(Fe_{0.973}Ru_{0.027})_2As_2$, before the paper end with summary of the results.

**Experimental**

High-quality single crystals of Ca(Fe$_{1-x}$Ru$_x$)$_2$As$_2$ were grown using the FeAs flux method. Mixtures with ratio of Ca : Fe : Ru : As = 1 : 4 − 4x : 4x : 4 were placed in alumina crucibles and sealed in Ta crucibles under argon atmosphere, heated to 1180℃ for 15 h and slowly cooled to 950℃ with a rate of 3 K/h before cooling to room temperature with a rate of 120 K/h. Mechanically isolated from the flux, plate like crystals were obtained with typical size 2 × 2 × 0.1 mm$^3$. A similar method was used for the growth of single crystals of Ca$_{1-y}$La$_y$(Fe$_{0.973}$Ru$_{0.027}$)$_2$As$_2$ with La content up y=0.2.

Previously the effect of post-growth annealing on air-quenched grown single crystals of CaFe$_2$As$_2$[18] and Ca(Fe$_{1-x}$Co$_x$)$_2$As$_2$[27,28] has been reported. Rapidly quenched crystals have substantial internal strain which has drastic influence on the physical properties. For our growth procedure with slow cooling only very little strain with negligible effect on the phase diagram is evidenced by the perfect similarity of the our phase diagram for Ca(Fe$_{1-x}$Ru$_x$)$_2$As$_2$ with that on CaFe$_2$As$_2$ under truly hydrostatic pressure[17].

The actual compositions x and y of the substitutions have been determined by energy dispersive X-ray fluorescence spectrometer (EDX) measurements. Three crystals were chosen for each concentration for the EDX measurements. Each crystal was investigated on two or three different positions. The crystals were found to be homogeneous with less than 5% relative variation of the actual compositions. All compositions listed in this paper are actual compositions which differ from the nominal (starting) compositions. All single crystals were characterized via X-ray diffraction with a Philips X'pert diffractometer using CuK$_{\alpha 1}$ radiation. The DC magnetic susceptibility was characterized using a superconducting quantum interference device magnetometer (Quantum Design, Inc.), whereas the electrical resistivity was measured using the standard four-probe method with a commercial (Quantum Design, Inc.) physical property measurement system (PPMS). The c-axis thermal expansion was measured by utilizing a high-resolution capacitive dilatometer in the PPMS.[29]

**Results and Discussions**

Fig. 1 shows the X-ray diffraction patterns of single crystalline Ca(Fe$_{1-x}$Ru$_x$)$_2$As$_2$ for $x$ = 0,0.015, 0.023, 0.027, and 0.065, respectively. Only (00l) peaks are observed, indicating the perfect orientation in the single crystals. Increasing the Ru doping level moves the peaks to higher diffraction angles. This corresponds to a shrinking of the c-axis as indicated in the inset of Fig. 1 and expected from previous studies of Ba(Fe$_{1-x}$Ru$_x$)$_2$As$_2$[30] and Ca(Fe$_{1-x}$Rh$_x$)$_2$As$_2$[20].

Fig. 2(a) displays the normalized electrical resistivity of single crystalline Ca(Fe$_{1-x}$Ru$_x$)$_2$As$_2$ (0≤x≤0.023) from 2K to 300 K. For x=0.023 a hysteresis between cooling and warming conditions is found. The resistive anomaly at 163 K for pure CaFe$_2$As$_2$ is associated with the coupled structural (Tet to Ort) and magnetic phase transitions. As Ru doping increases, the transition temperature is suppressed monotonically and the change of the resistivity at the transition evolves from a sharp feature in pure CaFe$_2$As$_2$ to a broadened signature, similar as found for Ba(Fe$_{1-x}$Ru$_x$)$_2$As$_2$[30]. As shown in the inset of Fig. 2(a), the transition temperature has been determined by the peak position in d(R(T))/dT to 139K and 116K for x=0.015 and 0.019 respectively. Note that the single peaks indicate joint structural and magnetic transitions. Contrary to the case of electron doping in Ca(Fe$_{1-x}$Co$_x$)$_2$As$_2$[31], the coupling of magnetic and structural transitions upon doping has previously been found in Ba(Fe$_{1-x}$Ru$_x$)$_2$As$_2$ and (Ba$_{1-x}$K$_x$)Fe$_2$As$_2$.[32, 33] As temperature decreases, the x=0.023 single crystal changes from the Tet phase to the Ort phase, and finally into the CT phase. The latter transition exhibits a significant hysteresis between cooling (42K) and warming (82K) conditions, indicating a first-order transition.[17] The observed resistivity fluctuations between the Ort and CT phase transitions likely reflect the competition of these two orderings upon cooling. Upon warming, due to the large hysteresis, there is no obvious indication of the Ort phase transition in the resistivity data. Furthermore, the resistivity displays a drop near 10 K but does not reach zero.

As shown in Fig. 2(b), with increasing Ru substitution for $0.023 \leq x \leq 0.065$ the Ort phase transition is completely suppressed and the CT phase is stabilized and shifted towards larger temperatures, i.e. 73 K (90 K) for x=0.027 and 165 K (175 K) for x=0.065 upon cooling (warming). Just like for $CaFe_2(As_{1-x}P_x)_2$[19] and $Ca(Fe_{1-x}Rh_x)_2As_2$[20], a change from non-Fermi-liquid to Fermi-liquid like electrical resistivity behavior is found in the CT phase for $Ca(Fe_{1-x}Ru_x)_2As_2$. In the lower inset of Fig. 2(b), it is displayed that the x=0.027 sample exhibits non Fermi liquid behavior, $\Delta R \propto T^{1.6}$, due to the antiferromagnetic fluctuation in the Tet phase [19, 20]. Respectively, the upper inset indicates Fermi liquid behavior, $\Delta R \propto T^2$ in the CT phase, similar as found previously, which could be related to the disappearance of magnetic fluctuations [19, 20]. The increase of the CT transition temperature with Ru doping shares similarities with that of $Ca(Fe_{1-x}Rh_x)_2As_2$[20] and $CaFe_2As_2$ under hydrostatic pressure[17].

Fig.3 shows high-field (5T) magnetic susceptibility data of $Ca(Fe_{1-x}Ru_x)_2As_2$ single crystal for $x = 0$, 0.015, and 0.027 for H//ab. In the high temperature Tet phase, an approximately linear relationship is observed in the temperature dependence of the magnetic susceptibility in all crystals. The $\chi(T)$ data for x=0 and 0.015 show clear SDW anomalies at transition temperatures 163K and 137K, consistent with the values determined form electrical resistivity. For x=0.027, the CT phase transition gives rise to a hysteresis between 75 K and 90 K upon cooling and warming again consistent with electrical resistivity and previous observations on $CaFe_2(As_{1-x}P_x)_2$[19] and $Ca(Fe_{1-x}Rh_x)_2As_2$[20] $(Ca_{1-x}Pr_x)Fe_2As_2$[21].

Based on these data we have constructed the temperature-composition phase diagram displayed in Fig. 4. At x=0.023 a discontinuous first-order change from the Ort to the CT ground state is observed. With further chemical pressure, the CT phase is stabilized. There is no bulk superconducting regime. The phase diagram resembles that found under truly hydrostatic pressure for $CaFe_2As_2$.[17] For 0.35 GPa hydrostatic pressure, $CaFe_2As_2$ also changes directly from the Ort phase into CT phase upon cooling, with a CT transition temperature of 40K, similar as found for x=0.023 at ambient pressure. Thus, similar to $Ba(Fe_{1-x}Ru_x)_2As_2$, partial Ru-substitution of the Fe atoms mainly acts as chemical pressure.

This together with the results of truly hydrostatic pressure[17] indicates that superconductivity in $CaFe_2As_2$ requires either anisotropic strain induced by non-hydrostatic pressure conditions or additional charge carrier doping. This is in contrast to the cases of $BaFe_2As_2$ and $SrFe_2As_2$[13]. The origin if this difference is, that for $CaFe_2As_2$ under hydrostatic or chemical pressure the CT ground-state appears (in a first-order transition) before the Ort/magnetic ground state if fully suppressed. Compared with isoelectronic Ru substitution, there is an additional electron doping effect in the case of $Ca(Fe_{1-x}Rh_x)_2As_2$[20]. For the latter series, the Ort/magnetic phase can be suppressed continuously down to low temperatures and a small superconductivity dome appears before entering the CT phase.[20]

In order to disentangle the effect of charge carrier doping from that of chemical pressure, we have chosen x=0.027 with well established CT ground state for a series with additional La-substitution of the Ca site, which gives rise to electron doping. Fig.5 shows the X-ray diffraction patterns of single crystalline $Ca_{1-y}La_y(Fe_{0.973}Ru_{0.027})_2As_2$ for $y = 0.04$, 0.06, 0.08, 0.15, and 0.20, respectively. Only (00l) peaks are observed, indicating good quality of the single crystals. From the inset of Fig.5, the (002) peaks of the diffraction patterns randomly vary from 15.21° to 15.26°. Concerning the (002) peak of $Ca(Fe_{0.973}Ru_{0.027})_2As_2$ located at 15.24°, these results imply that, with similar ionic radii as Ca, La doping does not obviously decreases the *c*-axis. Therefore, similar as for $Ca_{1-x}La_xFe_2As_2$[21], La substitution in

$Ca_{1-y}La_y(Fe_{0.973}Ru_{0.027})_2As_2$ realizes electron doping without additional chemical pressure.

At low La doping, i.e. y=0.04, the electrical resistivity and magnetic susceptibility data, cf. Fig. 6(a) and the lower inset of this figure, are rather similar to those of y=0. The first-order CT phase transition is slightly destabilized and observed at 64K upon cooling and 76 K during warming. Again Fermi liquid behavior is found within the CT phase (see upper inset). Further La doping, shifts the CT transition to lower values. As shown in Fig. 6(b) for y=0.06, a hysteresis is still found in magnetic susceptibility between 43K (cooling) and 49K (warming). On the other hand, the signature and hysteresis in the electrical resistivity (inset of Fig. 6(b)) is already very much suppressed similar as in $Ca_{1-x}Pr_xFe_2As_2$ for x=0.145[21]. Altogether the results indicate, that La doping in $Ca_{1-y}La_y(Fe_{0.973}Ru_{0.027})_2As_2$ decreases the CT transition temperature, the amplitude of the resistivity jump at the CT transition, and the width of hysteresis.

Indeed further increase of La doping completely suppresses the CT transition signatures in electrical resistivity and magnetic susceptibility, as shown for y=0.08 in Fig. 6(c). There is a broad kink around 90K in magnetic susceptibility in 5T (cf. upper inset of Fig.6(c)), which probably arises from the competition between the tetragonal and the CT phase. For this concentration two broad drops of the electrical resistivity at 42 K ($T_{c1}$) and 25 K ($T_{c2}$), which are easily suppressed by application of a small magnetic field (cf. lower inset of Fig.6(c)) may be associated with incipient superconductivity (not bulk, since neither complete suppression of the resistivity, nor significant diamagnetism occurs).

Further enhancement of the La-doping (y=0.15) leads to more pronounced superconducting anomalies in the electrical resistivity, which now at $T_{c2}$ approaches zero (lower inset of Fig. 6(d)). The upper transition ($T_{c1}$) signature becomes further enhanced at y=0.2 (Fig. 6(d) main part and upper inset). The low-field magnetic susceptibility for the latter La concentration y=0.2 also displays clear diamagnetic response associated with $T_{c1}$ and $T_{c2}$ both under zero-field cooling (ZFC) and field cooling (FC) conditions (Fig. 6(e)), which further evidences superconductivity. Note however, that the size of the diamagnetic signal is weak, corresponding to a superconducting volume fraction at low temperatures of order 3% only.

It is interesting to note, that two superconducting transitions were found in electrical resistivity measurements on $Ca_{1-x}La_xFe_2As_2$ with or without pressure[34, 35]. By contrast only one superconductivity transition has been found in $Sr_{1-x}La_xFe_2As_2$[36] and $Ba_{1-x}La_xFe_2As_2$ epitaxial thin films[37], with the transition temperature around 20 K, comparable with $T_{c2}$ in our system. Similar as in our system, for co-doped $(Ca_{1-y}La_y)Fe_2(As_{1-x}P_x)_2$ with x=0.06 and y=0.18 also an incomplete diamagnetic response has been found below $T_{c1}$ of order 40 K[38]. On the other hand for $Ca_{1-x}La_xFe_2As_2$ the upper transition at $T_{c1}$ is not seen in magnetic susceptibility indicating it is only very spurious. This comparison implies that Ru doping or chemical pressure stabilizes superconductivity above 40K in $Ca_{1-x}La_xFe_2As_2$.

Before summarizing the results on the La-doped material in a phase diagram, we provide final evidence for the complete suppression of the CT transition from c-axis thermal expansion (TE) data, obtained upon warming the sample in a capacitive dilatometer. As shown in Fig. 7, the clear length jump associated with the CT transition for y=0.06 is fully suppressed at y≥0.08, for which a smooth variation of the c-parameter is found. This proves a critical concentration $y_{cr}$ close to 0.08 for the CT transition in $Ca_{1-y}La_y(Fe_{0.973}Ru_{0.027})_2As_2$.

Fig. 8 displays the resulting phase diagram for the $Ca_{1-y}La_y(Fe_{0.973}Ru_{0.027})_2As_2$ series. Clearly with La-doping the CT phase can be suppressed completely. It has been discussed previously that the CT transition is driven by an enhancement of the interlayer As-As bonding [39]. Our individual control of chemical pressure and electron doping in $Ca(Fe_{1-x}Ru_x)_2As_2$ and $Ca_{1-y}La_y(Fe_{0.973}Ru_{0.027})_2As_2$ proves that these two parameters have different effects. While chemical pressure could stabilize the presence of interlayer As-As bonding, successive electron doping obviously weakens the bonding, and finally breaks it. So far, there are several examples, mentioned before, to induce the CT phase for $CaFe_2As_2$ based materials. However, our investigation represents the first example for tuning the CT phase back into the tetragonal phase.

Our results indicate that electron doping somehow counteracts chemical pressure in $CaFe_2As_2$. For $Ca_{1-x}Pr_xFe_2As_2$ the observed behavior therefore seems to be determined by the combination of both effects. While the CT transition is induced by chemical pressure, simultaneous electron doping decreases the strength of the interlayer As-As bonding and thereby suppresses the resistivity change at the CT transition.

Study of superconductivity has not been the main focus of this work, but we shortly comment on the observed signatures of incipient superconductivity when the CT phase is completely suppressed by electron doping. This suggests that the CT state is detrimental to superconductivity. According to inelastic neutron scattering and nuclear magnetic resonance measurements on strained $CaFe_2As_2$[40, 41], AF spin fluctuations and electronic correlations are completely absent in the CT state. For related $Ca_{1-x}Pr_xFe_2As_2$ with x=0.14, recent scanning transmission electron microscopy (STEM) and scanning tunneling microscopy (STM) found filamentary superconductivity related to an inhomogeneity of the Pr concentration on nanometer scale.[42] Respectively, incomplete (and possibly filamentary) superconductivity above 40 K in our $Ca_{1-y}La_y(Fe_{0.973}Ru_{0.027})_2As_2$ material for y=0.2 may also be related to small local variations of the La and Ru dopants which are impossible to resolve by our microprobe EDX experiments. Importantly, these local inhomogeneities could only lead to superconductivity if the CT phase is fully suppressed. A superconducting gap as large as 30 meV on localized regions indicate the intrinsic nature of the high temperature superconductivity in this system.[42] Compared to pure $(CaLa)Fe_2As_2$, chemical pressure in $(CaLa)(FeRu)_2As_2$ or $(CaLa)Fe_2(AsP)_2$ drives the material closer to the CT phase. The coinciding enhancement of the diamagnetic response at $T_{c1}$ suggests a stabilization of superconductivity under chemical pressure. This is in accordance with a recent theoretical scenario treating orbital fluctuations related to the CT transition as key ingredient to realize high $T_c$ values in these materials.[43]

## Conclusion

We have demonstrated that the partial substitution of Fe by Ru in $Ca(Fe_{1-x}Ru_x)_2As_2$ acts as chemical pressure. Near $x_{cr}$=0.023 a discontinuous transition between the AF Ort ground state and the CT phase with Fermi liquid properties without any superconducting dome was found. Starting within the CT state additional electron doping in $Ca_{1-y}La_y(Fe_{0.973}Ru_{0.027})_2As_2$ substantially weakens the interlayer As-As bonding and tunes the ground state back into the Tet phase. Superconductivity only appears within the Tet phase, after complete suppression of the CT phase. This highlights the importance of AF spin fluctuations and electronic correlations to the superconductivity in iron pnictide materials. Chemical pressure in $Ca_{1-x}La_xFe_2As_2$ enhances local superconductivity above 40K.

## Acknowledgements

The authors would like to thank Qinghua Zhang for the EDX measurement and Anton Jesche for helpful discussions. K. Z. thanks the Alexander von Humboldt foundation for financial support. Funding by DFG SPP1458 is acknowledged. Work at IOPCAS was supported by NSF & MOST of China through research projects.

# Figure Captions:

**Fig. 1.** X-ray diffraction pattern of single crystal Ca(Fe$_{1-x}$Ru$_x$)$_2$As$_2$ for $x$ = 0, 0.015, 0.023, 0.027, and 0.065, respectively. Inset: the $c$-axis lattice parameters derived from the (002) peak of the diffraction patterns

**Fig. 2(a).** Temperature-dependent normalized electrical resistivity of single crystal Ca(Fe$_{1-x}$Ru$_x$)$_2$As$_2$ (0<x<0.023), during cooling and warming conditions for x=0.023. (Lower inset) d$R(T)$/d$T$ versus $T$ from 80K to 210K for x=0, 0.015, and 0.019, respectively.

**Fig. 2(b).** Temperature-dependent normalized electrical resistivity of single crystal Ca(Fe$_{1-x}$Ru$_x$)$_2$As$_2$ (0.023<x<0.065). (Lower inset) ρ vs T$^{1.6}$ for Ca(Fe$_{1-x}$Ru$_x$)$_2$As$_2$ with x=0.027. (Upper inset) ρ vs. T$^2$ for Ca(Fe$_{1-x}$Ru$_x$)$_2$As$_2$ with x=0.027.

**Fig. 3.** Magnetic susceptibility $\chi(T)$ in a field of 5 T applied parallel to the ab plane of Ca(Fe$_{1-x}$Ru$_x$)$_2$As$_2$ for x=0, 0.015, and 0.027 respectively.

**Fig. 4.** Temperature-composition (T−x) phase diagram of single crystal Ca(Fe$_{1-x}$Ru$_x$)$_2$As$_2$ (0<x<0.065), $T_{SDW}$ represents the structural/magnetic transitions. $T_{cT}$ is the collapsed tetragonal transition temperatures on warming and cooling conditions.

**Fig. 5.** X-ray diffraction pattern of single crystal Ca$_{1-y}$La$_y$(Fe$_{0.973}$Ru$_{0.027}$)$_2$As$_2$ for $y$ = 0.04, 0.06, 0.08, 0.13, and 0.20, respectively. Inset: enlargement of the (002) peak of the diffraction patterns.

**Fig. 6(a).** Normalized electrical resistivity for single crystalline Ca$_{1-y}$La$_y$(Fe$_{0.973}$Ru$_{0.027}$)$_2$As$_2$ for $y$ = 0.04, during cooling and warming. (Upper inset) ρ vs. T$^2$ for $x$ = 0.04. (Lower inset) Magnetic susceptibility $\chi(T)$ for y=0.04, at cooling and warming conditions in *H//ab* under 5 T

**Fig. 6(b).** Magnetic susceptibility $\chi(T)$ for Ca$_{1-y}$La$_y$(Fe$_{0.973}$Ru$_{0.027}$)$_2$As$_2,$ y=0.06, measured at 5

T applied parallel to the ab plane. (Lower inset) $\rho(T)/\rho(80K)$ of single crystalline $Ca_{1-y}La_y(Fe_{0.973}Ru_{0.027})_2As_2$ for $y = 0.06$ during warming procedure.

**Fig. 6(c).** Normalized electrical resistivity of single crystal $Ca_{0.92}La_{0.08}(Fe_{0.973}Ru_{0.027})_2As_2$. (Lower inset) $\rho(T)/\rho(300K)$ versus $T$ for $y=0.08$ between 5K to 50K at several different magnetic fields. (Upper inset) $\chi(T)$ for $y=0.08$ in a field of 5 T applied parallel to the *ab* direction.

**Fig. 6(d).** Normalized electrical resistivity of single crystalline $Ca_{0.8}La_{0.2}(Fe_{0.973}Ru_{0.027})_2As_2$. (Upper inset) $\rho(T)/\rho(300\ K)$ versus $T$. (Lower inset) $\rho(T)/\rho(60K)$ versus $T$ of single crystal $Ca_{0.85}La_{0.15}(Fe_{0.973}Ru_{0.027})_2As_2$.

**Fig. 6(e).** Magnetic susceptibility $\chi(T)$ of $Ca_{0.8}La_{0.2}(Fe_{0.973}Ru_{0.027})_2As_2$ measured in a field of 30 Oe applied in ZFC and FC modes for $H//c$. (Lower inset) enlargement of ZFC and FC curves around two superconducting transitions at 41 K and 30 K.

**Fig. 7.** Temperature-dependent, c-axis relative length change $[\Delta c/c(300\ K)=[c(T)-c(300\ K)]/c(300\ K)]$ for single crystalline $Ca_{1-y}La_y(Fe_{0.973}Ru_{0.027})_2As_2$ for $y = 0.06$, 0.08, and 0.20. All data are obtained upon increasing the temperature.

**Fig. 8.** Phase diagram of single crystal $Ca_{1-y}La_y(Fe_{0.973}Ru_{0.027})_2As_2$ ($0<y<0.20$). $T_{cT}$ represents the collapsed tetragonal transition temperatures on warming and cooling conditions, $T_{c1}$ and $T_{c2}$ denote the two superconducting transitions in the tetragonal phase.

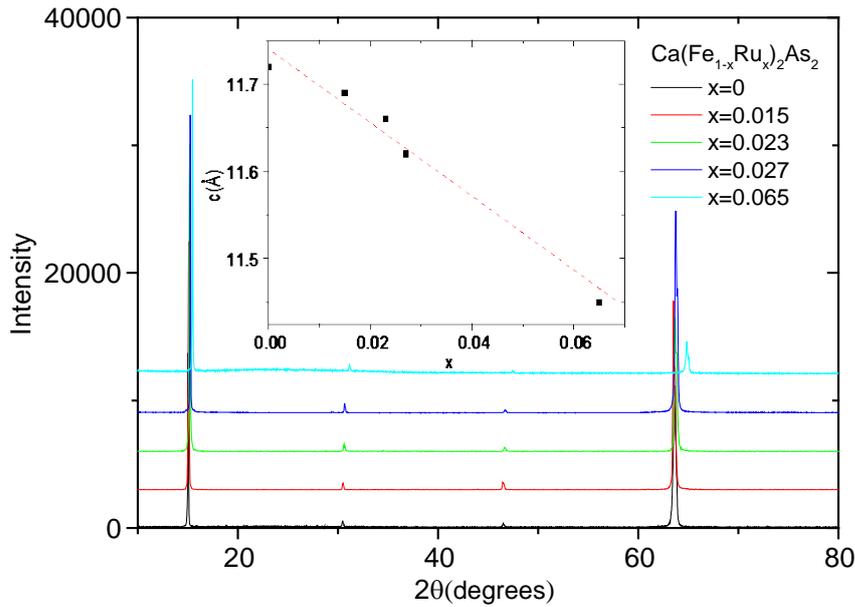

Fig. 1

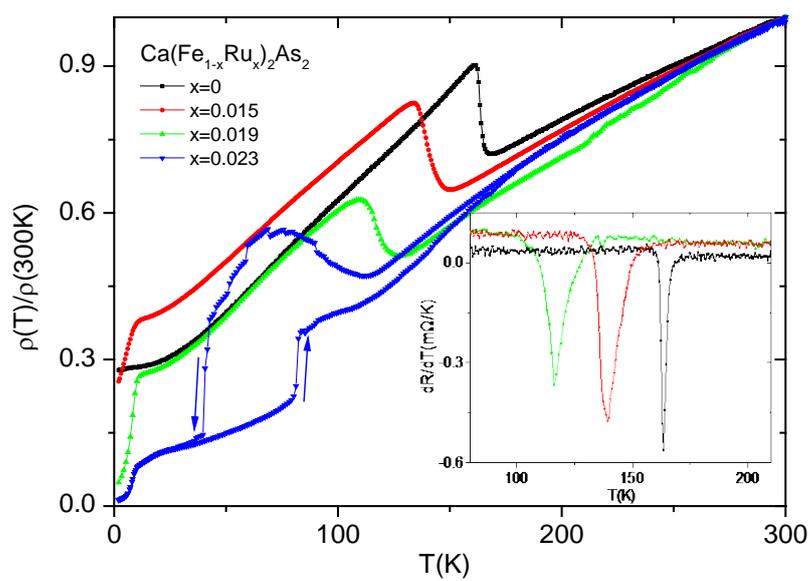

Fig. 2(a)

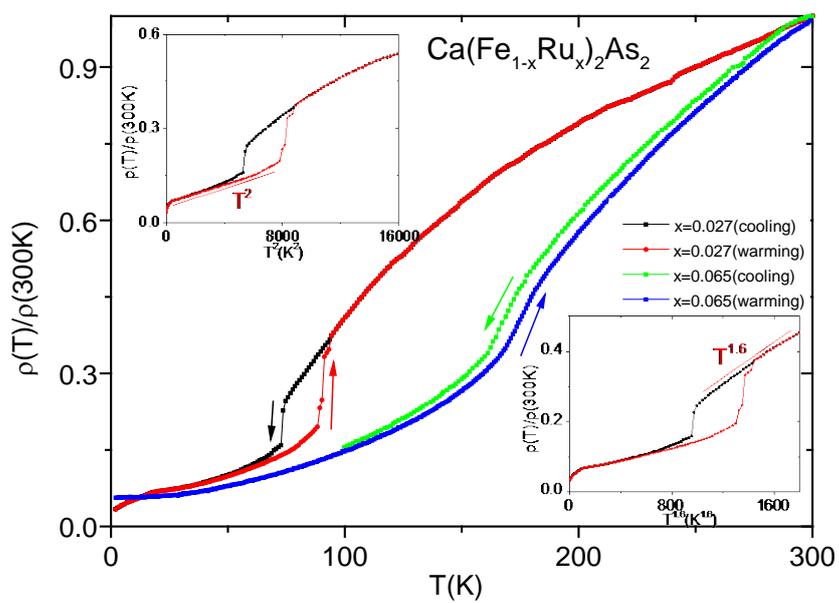

Fig. 2(b)

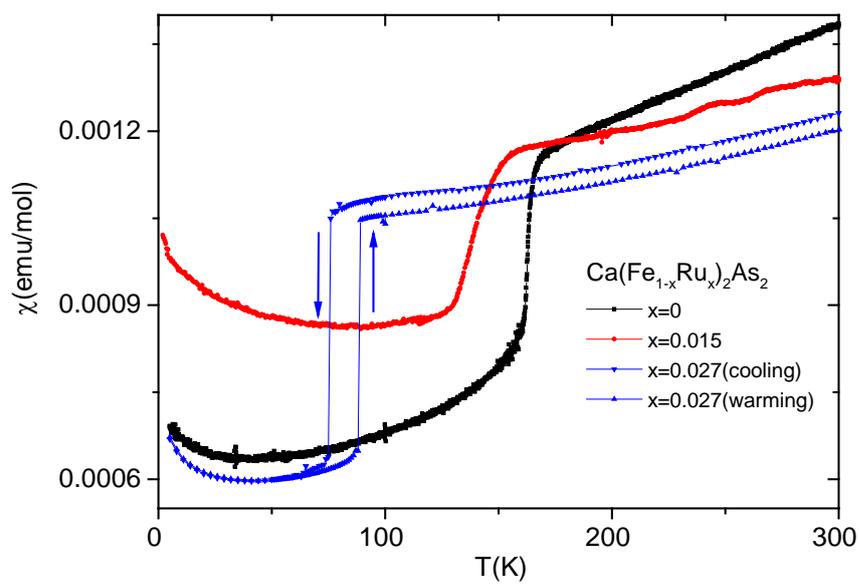

Fig. 3

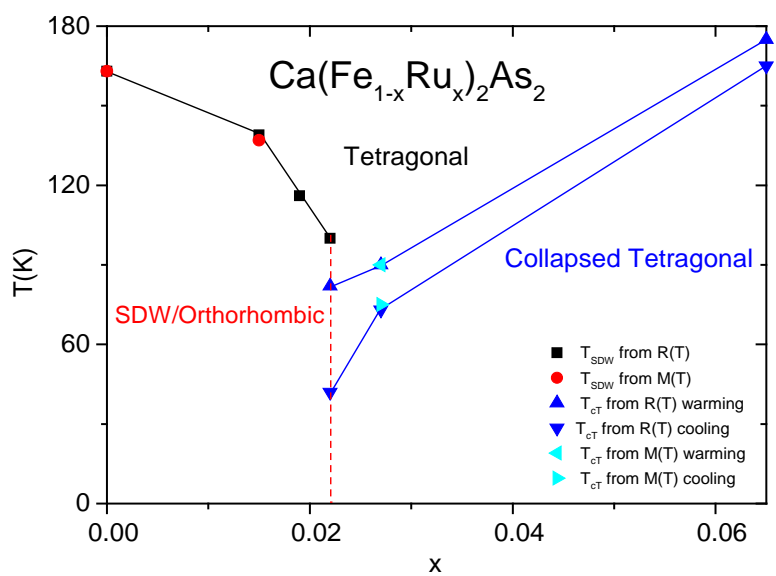

Fig. 4

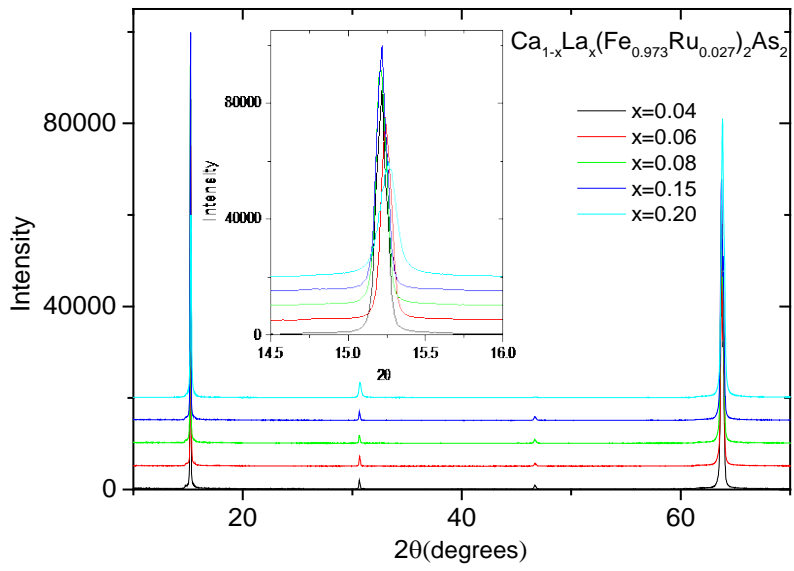

Fig. 5

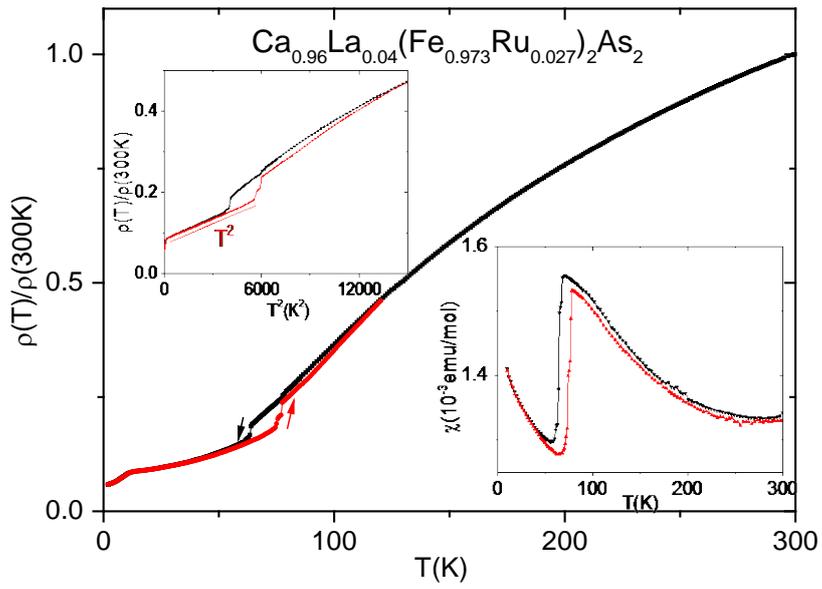

Fig. 6(a)

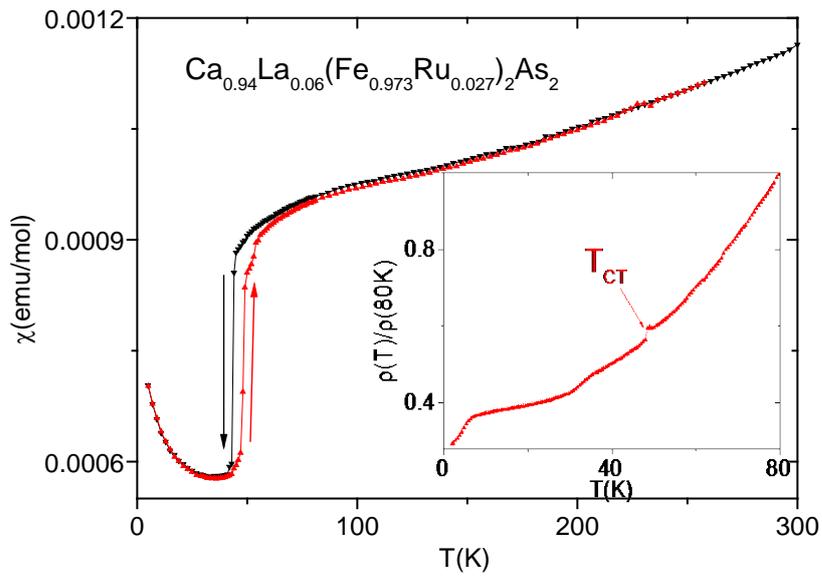

Fig. 6(b)

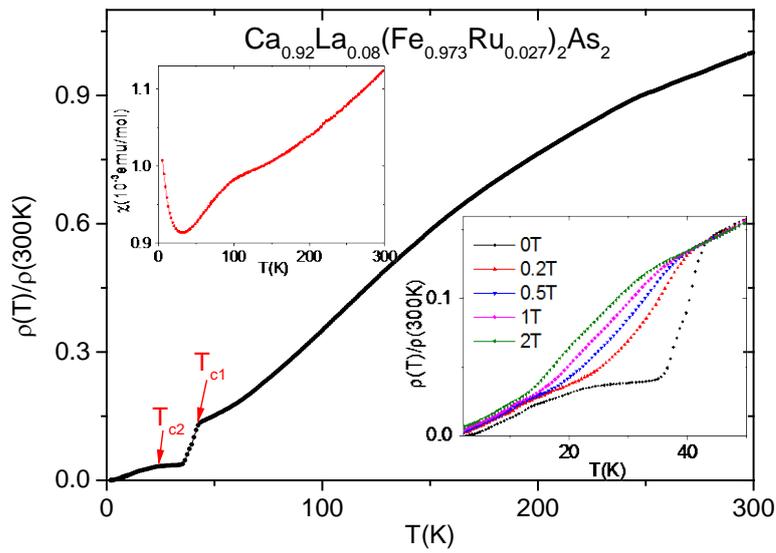

Fig. 6(c)

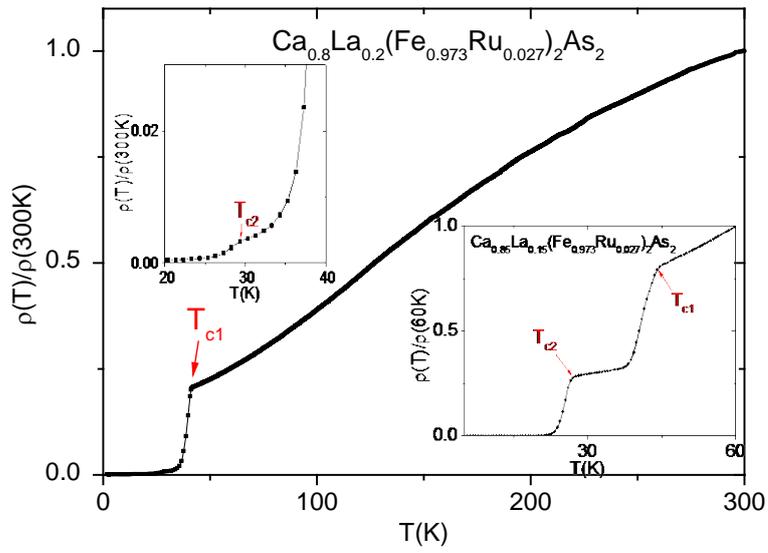

Fig. 6(d)

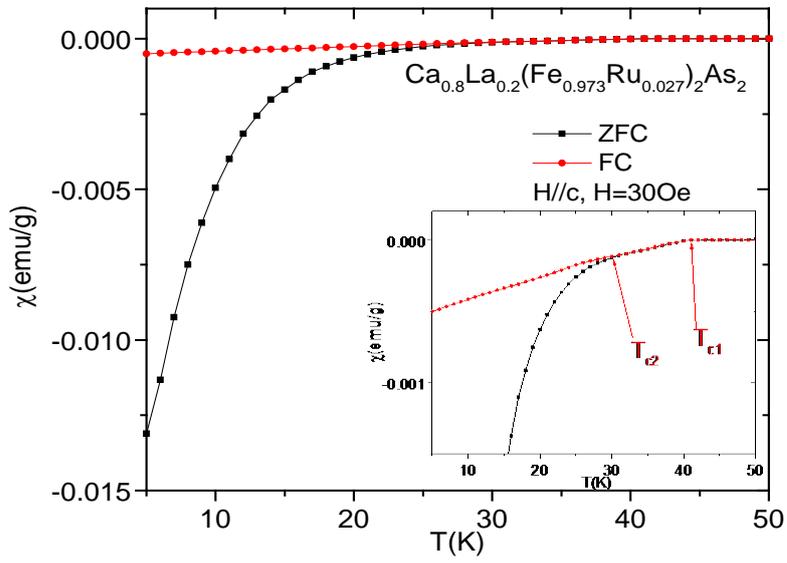

Fig. 6(e)

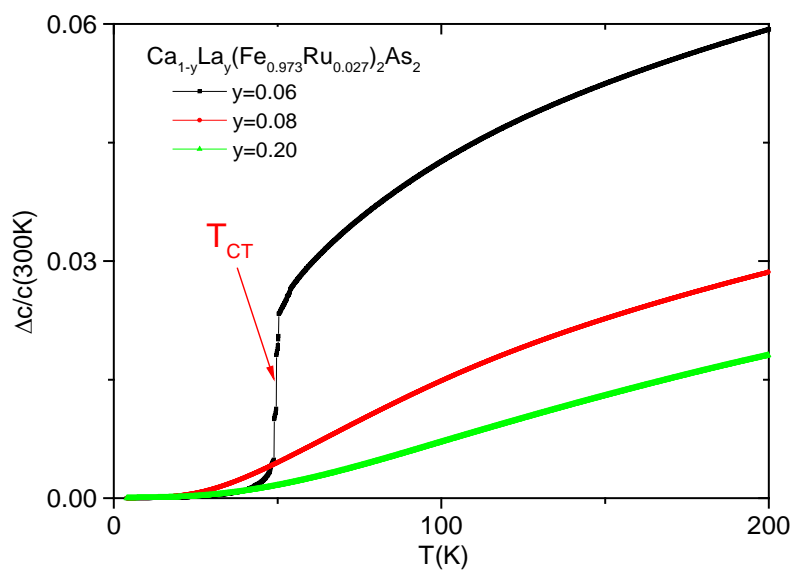

Fig. 7

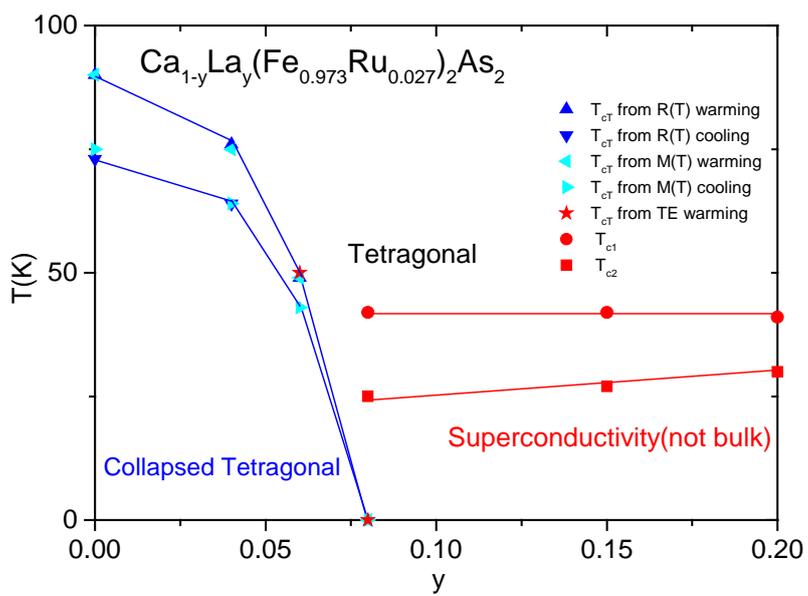

Fig. 8